\documentclass[preprint,aps]{revtex4}
\usepackage{mathrsfs}
\usepackage{amsmath}
\usepackage{graphicx}
\usepackage{multirow}
\usepackage{makecell}
\usepackage{booktabs}

\usepackage{dcolumn}
\usepackage{bm}
\usepackage{xcolor}
\usepackage[utf8]{inputenc}
\usepackage[T1]{fontenc}

\newcommand{\cvs}{CuV$_2$S$_4$}

\newcommand{\ef}{$E_{\rm F}$}

\newcommand{\yic}[1]{\textcolor{black}{#1}}
\newcommand{\yicc}[1]{\textcolor{black}{#1}}

\begin{document}


\title{Non-Fermi liquid behavior in a correlated flatband pyrochlore lattice}


\author{Jianwei Huang$^{1}$, Lei Chen$^{1,*}$, Yuefei Huang$^{2,*}$, Chandan Setty$^{1,*}$, Bin Gao$^{1}$, Yue Shi$^{3}$, Zhaoyu Liu$^{3}$, Yichen Zhang$^{1}$, Turgut Yilmaz$^{4}$, Elio Vescovo$^{4}$, Makoto Hashimoto$^{5}$, Donghui Lu$^{5}$, Boris I. Yakobson$^{2}$, Pengcheng Dai$^{1}$, Jiun-Haw Chu$^{3}$, Qimiao Si$^{1,\dagger}$, Ming Yi$^{1,\dagger}$}
\affiliation{
\\$^{1}$Department of Physics and Astronomy, Rice University, Houston, Texas 77005, USA
\\$^{2}$Department of Materials Science and NanoEngineering, Rice University, Houston, Texas 77005, USA
\\$^{3}$Department of Physics, University of Washington, Seattle, WA 98195, USA
\\$^{4}$National Synchrotron Light Source II, Brookhaven National Lab, Upton, New York 11973, USA
\\$^{5}$Stanford Synchrotron Radiation Lightsource, SLAC National Accelerator Laboratory, Menlo Park, California 94025, USA
\\$^{*}$ These authors contributed equally 
\\$^{\dagger}$To whom correspondence should be addressed: mingyi@rice.edu, qmsi@rice.edu
}

\date{\today}

\begin{abstract}
Electronic correlation effects are manifested in quantum materials when either the onsite Coulomb repulsion is large or the electron kinetic energy is small. The former is the dominant effect in the cuprate superconductors or heavy fermion systems while the latter in twisted bilayer graphene or geometrically frustrated metals. However, the simultaneous cooperation of both effects in the same quantum material--the design principle to produce a correlated topological flat bands pinned at the Fermi level--remains rare. Here, using angle-resolved photoemission spectroscopy, we report the observation of a flat band at the Fermi level in a 3$d$ pyrochlore metal \cvs. From a combination of first-principles calculations and slave-spin calculations, we understand the origin of this band to be a destructive quantum-interference effect associated with the V pyrochlore sublattice and further renormalization to the Fermi level by electron interactions in the partially filled V $t_{2g}$ orbitals. As a result, we find transport behavior that indicates a deviation from Fermi-liquid behavior as well as a large Sommerfeld coefficient. 
Our work demonstrates the pathway into correlated topology by constructing and pinning correlated flat bands near the Fermi level out of a pure $d$-electron system by the combined cooperation of local Coulomb interactions and geometric frustration in a pyrochlore lattice system.

\end{abstract}

\maketitle

\newpage

\section{Main}
Quantum many-body effects are manifested in materials where the electron kinetic energy ($t$) is small or comparable to the onsite Coulomb interactions ($U$)~\cite{Wigner1934}, $U/t\geq1$ leading to spontaneous symmetry breaking orders such as magnetism, nematicity, unconventional superconductivity, and charge density wave orders~\cite{Dagotto1994, Qazilbash2009a, Stewart1984, Gruner1988}. Such a regime can be reached either in materials with strong Coulomb interactions (large $U$) as the case in the high temperature superconducting cuprates, iron-based superconductors, or heavy fermion systems; or those with quenched kinetic energies (small $t$) by the construction of flat bands by moir\'e superlattice or destructive quantum interference~\cite{Bistritzer2011a, Cao2018a, Tang2011, Ye2021, Song2022, Ekahana2022}. In the former case, the electrons feel the strong repulsion from the nearby electrons and cannot be treated simply as single particles. \yic{As a result, the electron mass is often strongly enhanced and the band velocity strongly renormalized, which} often lead to non-Fermi liquid transport, quantum criticality, sometimes reaching the limit of Mott insulating phases~\cite{Dagotto1994, Imada1998, Gurvitch1987, Stewart1984, Si2001, Coleman2001, Senthil2004, Po2018}. 
In the latter case, in contrast to the conventional strong correlation system, a quasi-flat band with small bandwidth can be achieved for example in twisted-bilayer graphene by folding via the large moir\'e superlattice~\cite{Bistritzer2011a}. Similar strong correlation phenomena have also been found including unconventional superconductivity, ferromagnetism, linear resistivity~\cite{Cao2018a, Sharpe2019a, Polshyn2019}. 
Alternatively, another way to realize topological flat bands is via quantum interference of the electronic wavefunction on geometrically frustrated lattices, such as the kagome lattice consisting of corner-sharing triangles~\cite{Henley2010a, Tang2011, Regnault2022a, Calugaru2022, Neves2023}. In both twisted bilayer graphene and geometrically frustrated lattices, the non-local construction of the flat bands also deem these systems to be topological. In contrast to twisted bilayer graphene where access to devices are limited to certain experimental probes, geometrically frustrated lattices are found in a wide range of bulk materials for which large high quality single crystals are available~\cite{Liu2019, Ortiz2019b, Kang2020d, Yin2020}. However, while such quantum-interference-induced topological flat bands have been observed in the kagome materials, they still exhibit relatively large bandwidth and are often too far away from the Fermi level to be relevant for low energy physics~\cite{Li2018a, Kang2020d, Kang2020f}.

\begin{figure}
\includegraphics[width=0.98\textwidth]{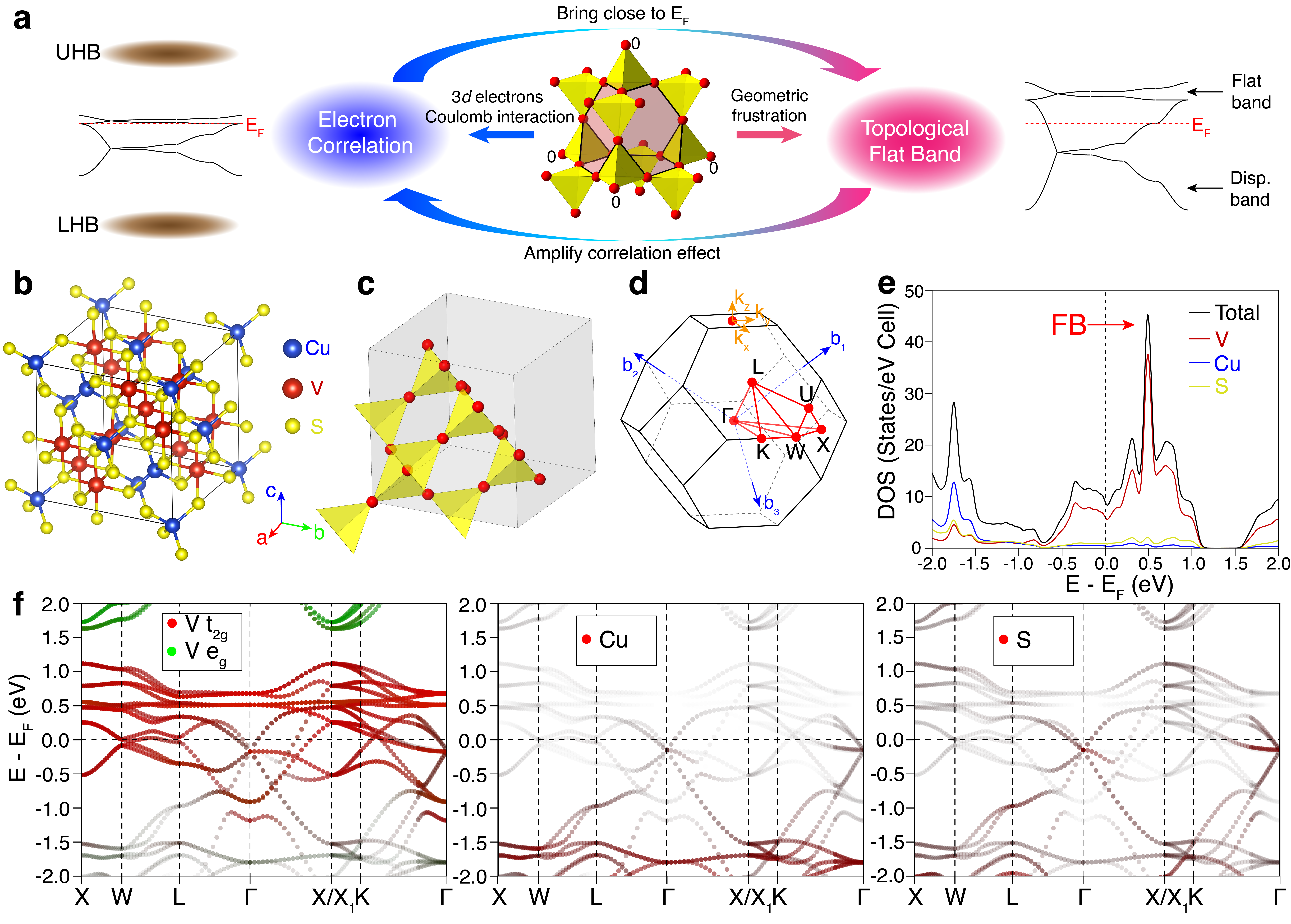}
\caption{\label{fig:Fig1}. {\bf Crystal structure and DFT calculations of~\cvs.} \textbf a, Correlation and 3D flat band in a pyrochlore lattice. The electrons are confined in the shaded region of the center of the pyrochlore lattice, with quenched effective hopping outside the region, leading to the 3D topological flat band. Moderate correlations of 3$d$ orbitals renormalize the 3D quantum interference induced topological flat band close to \ef, which in turn amplifies the correlation effects. \textbf b, Crystal structure of~\cvs. The V atoms are surrounded by the S tetrahedron and the Cu atoms are surrounded by the S octahedron. \textbf c, The V atoms form a pyrochlore sublattice which consists of corner-sharing tetrahedra. \textbf d, The 3D Brillouin zone (BZ), with the corresponding high symmetry points labeled. The blue arrows mark the primitive unit cell basis vectors. The yellow arrows mark the ARPES measurement coordinate. \textbf e, The density of states (DOS) and projected DOS by DFT calculations. The sharp peak located at 0.5 eV indicates the 3D flat band (FB). \textbf f, The band dispersion of \cvs~projected onto the V $t_{2g}$ and $e_g$ orbitals and Cu and S atoms by DFT calculations.
}
\end{figure}

An optimal way to further enhance $U/t$ is therefore to have both large $U$ and small $t$ by combining the two methods where electron-electron interactions are strong for a system that exhibits natural flat bands due to destructive quantum interference. In such a way, starting with quantum-interference induced topological flat bands from a single particle picture, additional Coulomb interactions can further enhance the effective mass and renormalize the flat bands towards the Fermi level, where interactions are more likely to lead to electronic instabilities.
\yic{Amongst the geometrically frustrated lattices, the kagome lattice has been intensely studied in recent years~\cite{Liu2019, Ortiz2019b, Kang2020d, Yin2020, Li2018a, Kang2020f}. While flat bands have been theoretically predicted and experimentally identified in kagome lattice materials, when realized in bulk, the finite inter-layer coupling disrupts the in-plane quantum interference effect, often leading to relatively large bandwidth. To find an optimal geometrically frustrated lattice, we seek materials that belong to the 3D analogue, the pyrochlore lattice. The pyrochlore lattice consists of corner-sharing tetrahedra, and provides destructive quantum-interference in 3D~\cite{Bergman2008, Guo2009, JHuang2023}. While numerous crystal structural families feature a pyrochlore sublattice, including traditional pyrochlores, Laves phases, and spinels, a metallic compound characterized by minimal hybridization between the electronic states arising from the constituents of the pyrochlore sublattice and those of the other atomic sites is particularly conducive to the formation of a flat band~\cite{Gardner2010}. In this context, the spinel \cvs~emerges as a highly promising candidate. Spinel compounds form with the general formulation of AB$_2$X$_4$ in a cubic crystal structure. The B sublattice, in particular, is a pyrochlore structure~\cite{Ghosh2019, Tsurkan2021}.}
By a combination of angle-resolved photoemission spectroscopy (ARPES), first-principles calculations, slave-spin calculations, transport and thermodynamic measurements, we identify a topological flat band near the Fermi level induced by a combined effect of destructive quantum interference due to a geometrically frustrated pyrochlore sublattice and on-site Coulomb interactions. Specifically, we find that the local Coulomb interactions renormalize the quantum-interference-induced topological flat band to near the Fermi level, which in turn further amplifies the electron correlations (Fig. 1a). This understanding is consistent with our observation of non-Fermi liquid transport behavior and large Sommerfeld coefficient in~\cvs.
Our study reveals a cooperative mechanism  between quantum-interference topological flat band and local Coulomb interactions, and lays the foundation for identifying model bulk systems for emergent phases in strongly correlated topological systems.
\cvs~is a face-centered cubic material \yic{that forms in the} space group of Fd\=3m (Fig.~\ref{fig:Fig1}b). The lattice parameter of the conventional unit cell is $a$ = 9.8 \AA~\cite{Hahn1956, LeNagard1979}. The Cu atoms form a diamond sublattice and the V atoms form a pyrochlore sublattice (Fig.~\ref{fig:Fig1}c). Importantly, the pyrochlore lattice is a geometrically frustrated lattice where three-dimensional (3D) destructive quantum interference confines the electron wavefunction to the center of the 3D block surrounded by the tetrahedra (Fig.~\ref{fig:Fig1}a). When only considering the nearest-neighbor hopping, the electrons have zero probability to travel outside the center block (Fig.~\ref{fig:Fig1}a), leading to 3D flat bands in momentum space~\cite{Bergman2008, Guo2009, JHuang2023}. This mechanism can perhaps be understood more intuitively by considering the pyrochlore lattice as alternately stacked kagome lattices and triangular lattices along the (111) direction with the interlayer coupling being completely quenched due to destructive quantum interference~\cite{Gardner2010, JHuang2023}. From our density-functional theory (DFT) calculations on \cvs~(Fig.~\ref{fig:Fig1}e and f), the electronic states near the Fermi level are dominated by the V 3$d$ orbitals while the contribution from Cu and S atoms are mainly 1 eV away from \ef. Notably, there is a sharp peak in the density of states (DOS) 0.5 eV above  \ef~(Fig.~\ref{fig:Fig1}e). This sharp peak indicates 3D quantum interference-induced flat bands originating from the V 3$d$ orbitals as further revealed by the calculated band dispersions (Fig.~\ref{fig:Fig1}f) ~\cite{Bergman2008, Guo2009, JHuang2023}. \yic{This calculation is consistent with that found independently in the materials database~\cite{Regnault2022a}. We can demonstrate the nature of this flat band to be from the quantum destructive interference effect by performing DFT calculations on an artificially distorted lattice of \cvs~after purposely disrupting the pyrochlore lattice by shifting two V atomic sites. The result shows the lack of such a sharp peak in the density of states and associated flat bands, therefore demonstrating that the flat band in the DFT calculation of \cvs~is indeed due to the quantum interference of the nearest neighbor hopping (see Supplementary Figure 10).} As the low energy electronic states are dominated by the V atoms with negligible hybridization to the Cu and S states, the pyrochlore physics is expected to be manifested in this compound, leading to flat bands that are topologically nontrivial~\cite{Guo2009, JHuang2023}. We note that previous studies have revealed two charge density wave transitions in \cvs~at around 50 K and 90 K with an accompanied structural transition from cubic to orthorhombic~\cite{Fleming1981a, Okada2004, Kawaguchi2012}, which are also consistently reproduced by our transport studies (Supplementary Figure 1). However, the lattice distortion is very small (less than 0.05\% for $a$, $b$ and less than 0.2\% for $c$~\cite{Kawaguchi2012}) and we do not find any significant modification of the electronic structure from both ARPES measurements and DFT calculations (Supplementary Note 2). Therefore for simplicity, we focus our discussions on the cubic structure of the compound for the rest of the paper, which will not affect our conclusions.

\begin{figure}
\includegraphics[width=0.98\textwidth]{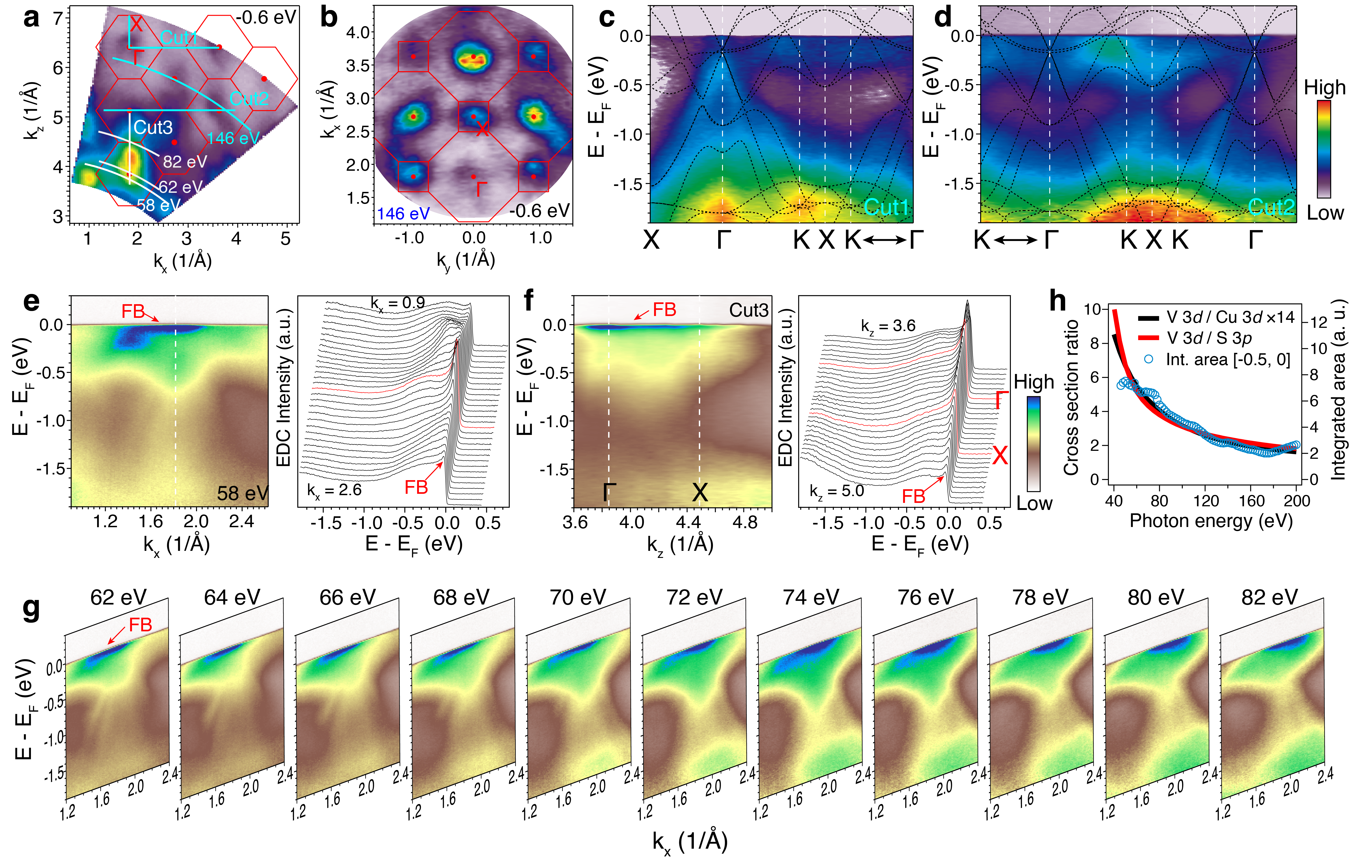}
\caption{\label{fig:Fig2}. {\bf The electronic structure and 3D flat band of \cvs~by ARPES.} \textbf a, Out-of-plane constant-energy contour mapping at -0.6 eV measured by varying different photon energies with respect to the (001) surface. The red solid lines mark the corresponding BZ. The cyan and white solid lines mark the cut positions. \textbf b, In-plane  constant-energy contour mapping at -0.6 eV of the (001) surface. \textbf c,\textbf d, Spectral images of cut1 and cut2 indicated in (\textbf a) with the corresponding band dispersion by DFT calculations (black dashed lines) overlaid on top. \textbf e, Spectral images and corresponding energy distribution curve (EDC) stacks showing the flat band around  \ef~measured by 58 eV photons with the corresponding momentum position indicated in (\textbf a). \textbf f, Spectral images and corresponding EDC stacks showing the flat band around  \ef~along the out-of-plane direction as indicated in (\textbf a). \textbf g, Spectral images measured with different photon energies showing the 3D flat band. The corresponding cut positions are indicated in (\textbf a). \textbf h, Photoionization cross-section ratio of V 3$d$ over Cu 3$d$ and S 3$p$ oritals (https://vuo.elettra.eu/services/elements/WebElements.html) and integrated spectral weight of the EDCs in (\textbf f) over the energy range of -0.5 $\sim$ 0 eV as a function of the photon energy.
}
\end{figure}

Having revealed theoretically the existence of 3D flat bands by geometric frustration, we then present the ARPES results. Figure \ref{fig:Fig2}a shows an out-of-plane constant energy contour (CEC) mapping of \cvs~at -0.6 eV with respect to the (001) surface obtained by varying the photon energy. The corresponding in-plane CEC mapping at -0.6 eV is shown in Fig.~\ref{fig:Fig2}b using 146 eV photons. Both mappings exhibit a spectral intensity matching well with the Brillouin zone (BZ) periodicity. From these, we are able to obtain precise cuts along the high symmetry momentum directions and compare them to the DFT calculations (Fig.~\ref{fig:Fig2}c and d). Overall, the measured spectral images match well with the calculated band dispersion below -0.5 eV, where the density of states is largely dominated by Cu and S orbitals, which is consistent with a previous photoemission study~\cite{Matsuno2001}. Near \ef, however, the measured electronic structure exhibits moderate deviations from DFT calculations. Surprisingly, we observe a flat band near the Fermi level in the spectral image. Its intensity is more strongly manifested when measured at smaller photon energies. An example of this is shown along the momentum cut indicated in Fig.~\ref{fig:Fig2}a as measured by 58 eV photons (Fig.~\ref{fig:Fig2}e). To examine the flat band around \ef~more closely, we plot the stack of energy distribution curves (EDC) for the corresponding spectral image. The sharp peaks in the EDCs marked by the red arrow clearly confirm the flat band around \ef. Next, we examine this flat feature along the out-of-plane direction (Fig.~\ref{fig:Fig2}f), where the sharp peaks are also observed at \ef~in the corresponding EDC stack. This can be further illustrated by a series of spectral images measured with different photon energies, where the flat band is observed to span almost the entire BZ, thus revealing its 3D nature (Fig.~\ref{fig:Fig2}g). Moreover, the observation of the stronger flat band intensity with relatively lower photon energies is consistent with it having dominantly V 3$d$ orbital character, as the expected photoionization cross-section ratio of the V 3$d$ over Cu 3$d$ and S 3$p$ orbitals (https://vuo.elettra.eu/services/elements/WebElements.html) increases as a function of lowering photon energy in the range of photon energies that we experimentally probed (Fig.~\ref{fig:Fig2}h).

\begin{figure}
\includegraphics[width=0.98\textwidth]{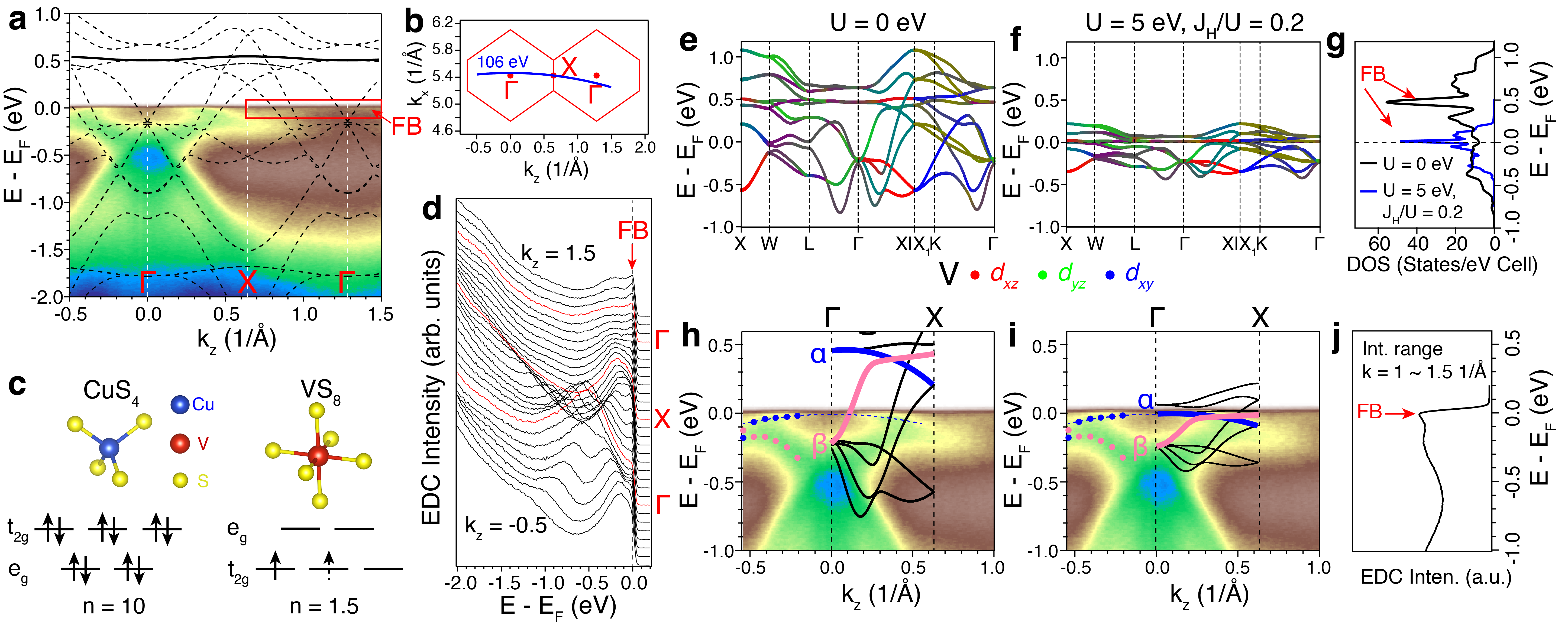}
\caption{\label{fig:Fig3}. {\bf Electron correlation effects and the flat band at \ef.} \textbf a, Spectral images measured on the (110) surface with the corresponding DFT calculations overlaid. Data taken at 106 eV. The solid line is the flat band at around 0.5 eV by calculations. The red arrow points to the measured flat band around \ef. \textbf b, Brillouin zone and the momentum cut position of (\textbf a). \textbf c, The crystal field splitting and electron filling of 3$d$ orbitals of Cu and V atoms. The Cu 3$d$ orbitals are completely filled by 10 electrons. The V three $t_{2g}$ orbitals are filled by an average of 1.5 electrons. \textbf d, EDC stacks of the spectral images in (\textbf a). The red arrow marks the flat band around \ef. \textbf e, The band dispersion of 12-band model by slave-spin calculations without including electron correlations ($U$ = 0 eV, $J_H$ = 0 eV). \textbf f, Same as (\textbf e) but including electron correlations ($U$ = 5 eV, $J_H$/$U$ = 0.2). \textbf g, DOS of band dispersion by the slave-spin calculations. \textbf h, Comparison of the spectral image by ARPES and slave-spin calculations (overlaid solid lines) with U = 0 eV along $\Gamma$ - X direction. The blue and pink solid circles are fitted band dispersions, which match with the blue ($\alpha$) and pink ($\beta$) solid lines from the calculated bands. The blue dashed line is a quadratic fitting of the $\alpha$ band (blue solid circles) that gives a band top of -7 meV. \textbf i, Same as (\textbf h) but with $U$ = 5 eV, $J_H$/$U$ = 0.2. \textbf j, The integrated EDC of the spectral image in (\textbf a) showing the flat band peak around \ef.
}
\end{figure}

While both our DFT calculations and ARPES measurements have revealed the existence of 3D flat band in~\cvs, the energy position of the flat band between them is different. The flat band predicted by the DFT calculations is 0.5 eV above the Fermi level while it is observed to be located at \ef~experimentally (Fig.~\ref{fig:Fig3}a). This disagreement is not only limited to the flat band, but also other bands in the near \ef~region, suggesting that the V 3$d$-dominated bands may exhibit non-negligible electron correlation effects. To understand the origin of the correlation effects, we note that in~\cvs, each Cu atom is surrounded by S atoms forming a tetrahetron while each V atom on the pyrochlore sublattice is surrounded by S atoms to form an octahedron (Fig.~\ref{fig:Fig3}c). As a result, the 3$d$ orbitals of Cu are split into $t_{2g}$ and $e_g$ orbitals by the crystal field with the $t_{2g}$ orbitals located at a higher energy. The situation is reversed regarding the V 3$d$ orbitals in the octahedral environment where the $e_g$ orbitals sit higher in energy. From X-ray photoemission spectroscopy (XPS) and X-ray emission spectroscopy (XES) measurements, it has been reported that the Cu 3$d$ orbitals are fully occupied with 10 valence electrons~\cite{Lu1996}. This is consistent with the negligible Cu 3$d$ states near \ef~from DFT calculations (Fig.~\ref{fig:Fig1}e and f). For the V 3$d$ orbitals, an average of 1.5 valence electrons fill the 3 degenerate $t_{2g}$ orbitals, where electron correlation effects are expected with suppression of coherent quasiparticle spectral weight $Z$~\cite{Georges2013a}. This distinction between the V and Cu atomic environments is consistent with the observation that the fully occupied Cu $d$ bands far away from \ef~are not strongly renormalized and agree well with the DFT calculations while those for the near-\ef~V $d$ orbitals show a mismatch between DFT and measured dispersions (Fig.~\ref{fig:Fig2}c, d, \ref{fig:Fig3}a and Supplementary Figure 5). This strong orbital-dependent \yic{renormalization} here in \cvs~is reminiscent of the orbital-selectivity in the iron-based superconductors where electron correlation effects are strongly enhanced for the Fe $d_{xy}$ orbital compared to other 3$d$ orbitals~\cite{Georges2013a, Misawa2012, Yu2013b, Yi2013a, Yi2015b, Yi2017c, Huang2022}.

To incorporate the electron correlation effects to the near-\ef~states in the calculated electronic structure, we performed a slave-spin calculation of a 12-band model comprised of the V $t_{2g}$ orbitals on the pyrochlore sublattice obtained by fitting the DFT results of \cvs~\yic{in the near \ef~region} (details of the calculations can be found in method and Supplementary Note 5). The calculated bands without incorporating the Coulomb interactions are shown in Fig.~\ref{fig:Fig3}e, which clearly reproduce the DFT calculations of the V 3$d$ bands near \ef. Specifically, the 3D flat band is 0.5 eV above \ef. \yic{As has been done for the iron-chalcogenide superconductors, we consider the effect of electron correlations by including both Coulomb interactions ($U$) and Hund's coupling ($J_H$). At a combination of $U$ = 5 eV and $J_H$/$U$ = 0.2}, the overall band structure exhibits a moderate band renormalization (Fig.~\ref{fig:Fig3}f). More interestingly, the 3D flat band is shifted close to the Fermi level, as can also be clearly seen in the shift of the sharp peak in the calculated DOS to near \ef~(Fig.~\ref{fig:Fig3}f and g). This is due to the particle-hole asymmetry and a charge transfer from coherent to incoherent part when we add interactions~\cite{Huang2022}. Indeed, a comparison of the two calculations with the corresponding ARPES spectral image along the $\Gamma$--X direction shows a much improved agreement for the one incorporating correlation effects (Fig.~\ref{fig:Fig3}h and i). Importantly, this improvement is not limited to the flat band around \ef, but also for the dispersive low energy electronic states around the $\Gamma$ and X points. To examine better the comparison of the calculations and the dispersive bands near \ef, we use a high symmetry cut taken with a higher photon energy of 106 eV, where the relative intensity of the V bands to Cu/S is lower. The dispersive bands, in particular the hole-like $\alpha$ band and the electron-like $\beta$ band, while dominantly of V $t_{2g}$ character, has a slightly higher hybridized contribution from Cu/S orbitals than the flat bands (Fig.~\ref{fig:Fig1}f, \ref{fig:Fig3}h and \ref{fig:Fig3}i), and hence more observable. We plot the band positions obtained from fitting EDCs in pink and blue markers in Fig.~\ref{fig:Fig3}h-i, respectively. We can see that their band velocities are much renormalized compared to the calculations at $U$ = 0 eV, but both match much better with \yic{the calculations at $U$ = 5 eV and $J_H$/$U$ = 0.2}. Moreover, the band top of the $\alpha$ band meets the topological flat band in a quadratic band touching point at the $\Gamma$ point, which is a well-known property for the pyrochlore lattice~\cite{Sun2009}. Taking advantage of this, we can estimate the flat band energy position by fitting the $\alpha$ band dispersion from experiments. The fitted band top is at -7 meV (Fig.~\ref{fig:Fig3}h-i). Considering a tiny gap (7 meV from DFT) opening at the $\Gamma$ point in the orthorhombic phase (see Supplementary Figure 2), the flat band should be in proximity of \ef. Since the energy of the flat band at $\Gamma$ is also near its band bottom (Fig.~\ref{fig:Fig3}a), we conclude that the flat band spectral peaks we observe around \ef~is likely the tail of its spectral intensity cut off by the Fermi Dirac function \yic{with the true energy position slightly above \ef~(Fig.~\ref{fig:Fig2}g). This is consistent with a cut measured at high temperature divided by Fermi-Dirac function (see Supplementary Note 8 and Supplementary Figure 14).} Hence the consistently observed renormalization of these dispersive bands near \ef~as well as the presence of the flat band at \ef~further demonstrates that the origin of this renormalizaiton is the electron correlation associated with the partially filled pyrochlore V $t_{2g}$ orbitals. \yic{We emphasize here that the existence of the flat band at \ef~must be a combined effect of electron correlations and destructive interference. Slave-spin calculations from the starting point of the distorted \cvs~crystal structure with destroyed pyrochlore sublattice does not produce a similar flat band at \ef, hence confirming that electron correlations alone are not sufficient to give rise to the flat band in \cvs~(see Supplementary Note 6).}

\begin{figure}
\includegraphics[width=0.8\textwidth]{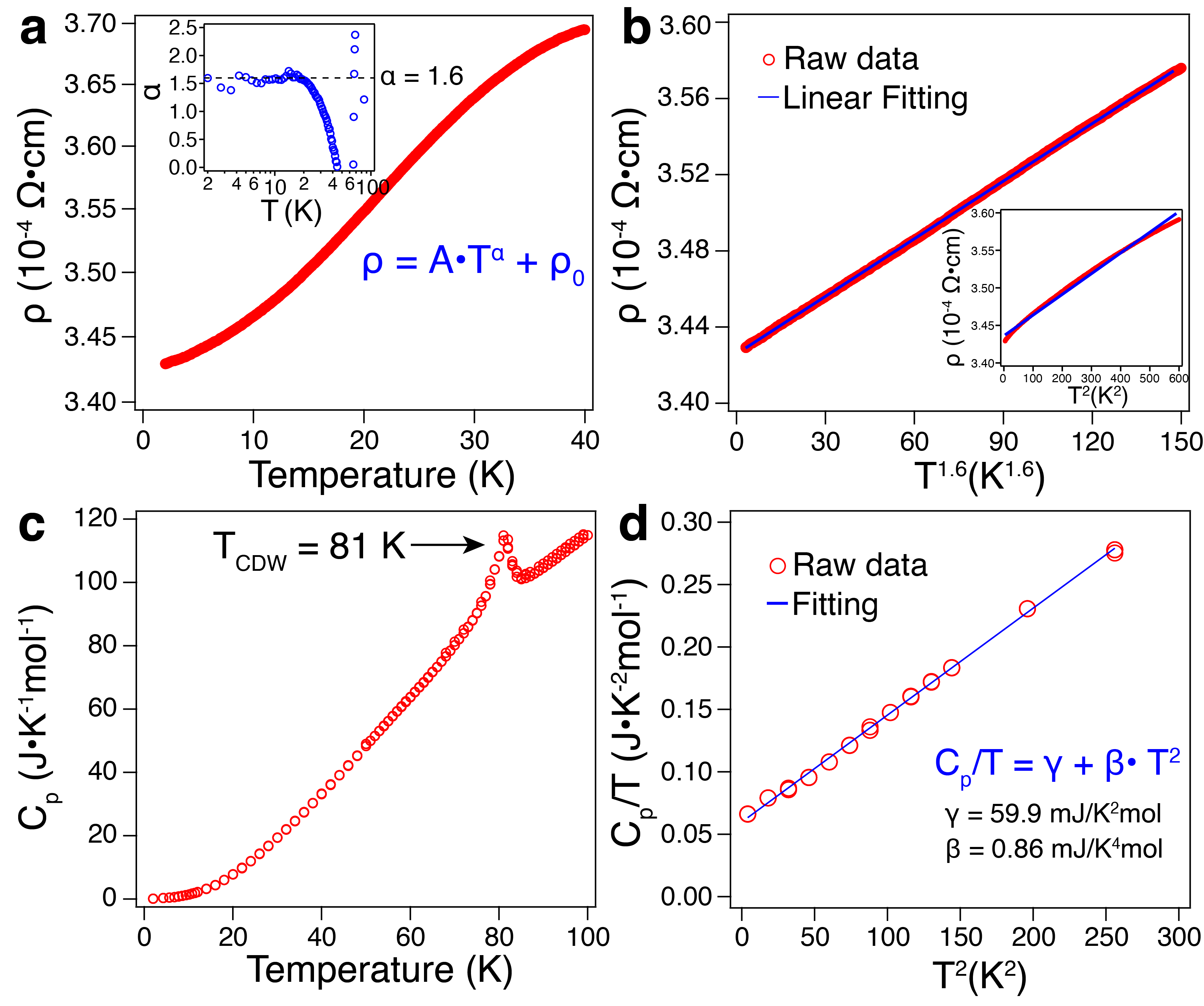}
\caption{\label{fig:Fig4}. {\bf Transport and thermodynamic measurements showing the non-Fermi liquid behavior.} \textbf a, Resistivity of \cvs. The inset shows the power law exponent $\alpha$ of the resistivity as a function of temperature. The resistivity has a power law behavior with $\alpha$ = 1.6 up to 20 K. \textbf b, Low temperature resistivity plotted as a function of $T^{1.6}$ and a linear fit. The inset shows resistivity plot as a function of $T^2$ and a linear fit (blue solid line). \textbf c, Heat capacity measurement of~\cvs. The black arrow points to a charge density wave transition at around 81 K. \textbf d, Linear fitting of C$_p$/$T$ as a function of $T^2$ to obtain the Sommerfeld coefficient $\gamma$. 
}
\end{figure}

The combination of ARPES and slave-spin calculations suggest that \cvs~is in a regime with large $U/t$, a result of both large $U$ from on-site Coulomb interaction and small $t$ from quantum interference effect associated with the pyrochlore lattice. For materials in the large $U/t$ regime, such as the unconventional superconductors and heavy fermion compounds, transport behavior often deviates from Fermi liquid description~\cite{Gurvitch1987, Butch2012}. We therefore examined the transport properties of \cvs. Interestingly, the temperature-dependent resistivity shows a power law behavior, $\rho(T)\sim T^\alpha$. To extract the exponent $\alpha$, we plot $\alpha = \partial \ln (\rho (T)-\rho_0)/ \partial \ln (T)$ as a function of temperature (inset of Fig.~\ref{fig:Fig4}a), which shows that $\alpha$ = 1.6 in the low temperature regime extending from the lowest temperature of 2 K up to 20 K. This is further evident in the much more linear behavior of the resistivity plotted as a function of $T^{1.6}$ compared to the plot as a function of $T^2$ (Fig.~\ref{fig:Fig4}b). This $T^{1.6}$ power law behavior deviates from the $\alpha$ = 2 behavior expected of Fermi liquids, \yicc{suggesting non-Fermi liquid behavior in \cvs. This is further indicated by the non-saturating magnetic susceptibility which follows logarithmic temperature variation at low temperatures (Supplementary Note 7 and Supplementary Figure 12).} 
Moreover, we also carried out specific heat measurements, from which we extract a Sommerfeld coefficient, $\gamma$, to be 60 mJ/K$^2$mol (Fig.~\ref{fig:Fig4}d). Notably, this $\gamma$ is 6 times larger than that predicted by the DFT calculations ($\gamma_{DFT}$=10 mJ/K$^2$mol), consistent with previous reports~\cite{Hagino1994, Gauzzi2019}. Given that this value is experimentally extracted in the CDW ordered state where the density of states develop CDW gap (Fig.~\ref{fig:Fig4}c), this enhancement factor is likely only a lower bound of the true factor and indicates the contribution from electron-electron correlations and the renormalized flat band near \ef. With that said, we also note that specific heat may not be as good a measure of the non-Fermi liquid behavior as transport also due to the potential gapping of the Fermi surface in the CDW phase, while resistivity derives from the remnants of the Fermi surface and its temperature dependence captures the non-Fermi liquid property.

We now advance a mechanism of the non-Fermi liquid behavior based on magnetic frustration. Intuitively, flat bands could be represented in real space by molecular orbitals that are essentially localized. Such a mapping has recently been advanced in a simpler variant of the lattice that features a flat band~\cite{chen2023metallic, Chen2022, Hu2023}. The topologically nontrivial flat bands need to be combined with other wide bands to be represented by exponentially localized symmetry-preserving (Kramers-doublet) Wannier orbitals, with essentially localized Wannier states that predominantly come from the flat bands and more extended Wannier states that are primarily associated with the wide bands. This leads to a Kondo-lattice description, with the two sets of Wannier states acting as the analogue of the $f$- and $spd$-orbitals of the heavy fermion materials. Importantly, in our pyrochlore case, the analogue-$f$-moments are geometrically frustrated. This means that the mapped Kondo lattice model is in the highly-quantum-fluctuating (large $G$) regime of the global phase diagram of the Kondo lattice~\cite{Paschen2020}, where the analogue-$f$-moments can form a spin liquid and the corresponding metallic phase can be a non-Fermi liquid phase. It has been proposed that such a phase is realized in the geometrically-frustrated heavy fermion material CePdAl~\cite{Zhao2019}, which displays non-Fermi liquid temperature dependence  in the electrical resistivity $\rho \sim T^{\alpha}$ with an exponent $\alpha$ that is smaller than the Fermi liquid value $2$. We propose that the same phase underlies the non-Fermi liquid behavior we have observed here in \cvs.

Hence from the combination of ARPES, DFT, slave-spin calculations, transport and heat capacity measurements, we have come to understand the \cvs~as a system where non-Fermi liquid behavior emerges from a cooperative result of quantum interference-driven quenching of kinetic energy through the V pyrochlore sublattice together with the Coulomb interactions of the V 3$d$ $t_{2g}$ orbitals. The spectral evidence of this is a topological flat band that forms out of geometric frustration in the single particle picture that is then renormalized and pinned to the Fermi level by Coulomb interactions. \cvs~is a beautiful example of the design principle of creating strongly correlated topological flat bands purely out of a $d$-electron system by the cooperative intertwinement of reducing $t$ and increasing $U$. It is important to point out that neither effect by itself can create the observed outcome: the destructive quantum interference by itself in \cvs~produces a flat band that is too far away from \ef~to affect transport behavior; the moderate Coulomb interactions here by itself can renormalize dispersive bands to an extend but cannot bring a large DOS to \ef. However, the renormalization of the topological flat band amplifies the correlation effects by renormalizing a large DOS to \ef, pushing the system towards the regime where emergent phases can arise from the large degeneracy produced in the vicinity of \ef. In a larger scope, \cvs~belongs to a wider iso-structural family of spinel 124 compounds where a pyrochlore sublattice can be found. LiV$_2$O$_4$ is another notable example where strong correlation effects have been reported, where the same V pyrochlore sublattice is found~\cite{kondo1997}. 
Going forward, the spinel 124 structural family provides a rich material platform for systematic exploration of further tuning of both Coulomb interactions and quantum interference effects by charge carrier and/or magnetic doping. The topological nature of the bands further implies the existence of surface states that could be observed in future studies. In an even broader context, the pyrochlore lattice is one of the many geometrically frustrated lattices that host topological flat bands. Our work lays out a pathway where local Coulomb interactions and quantum interference effects can be combined to amplify correlation effects. For such constructs, there still exists a vast material base to be explored.



\section{Methods}
\subsection{Sample growth and characterization}
Polycrystalline \cvs~was first prepared by a solid state method. The stoichiometric mixture of high purity Cu (99.99\%), V (99.9\%) and S (99.9\%) powders was ground and pressed into a pellet inside an argon glove box. The pellet was sealed in an evacuated quartz tube and heated at 1123 K for one week. The pellet was then ground again, sealed in an evacuated quartz tube together with iodine as the transport agent. The tube was placed in a tube furnace for two weeks, with the hot end held at 1123 K and the cold end at 1023 K. Crystals with an maximum size of $1\times 1\times1$ mm$^3$ were found in the cold end.

\subsection{ARPES measurements}
ARPES experiments were performed at beamlines 5-2 of the Stanford Synchrotron Radiation Lightsource and NSLS-II ESM beamlines of Brookhaven National Laboratory. Both beamlines are equipped with DA30 electron analyzers. Results were reproduced at both facilities. The angular resolution was set to 0.3$^\circ$. The total energy resolution was set to 20 meV or better. All the samples were cleaved \textit{in-situ} at 15 K and all the measurements were conducted in ultra-high vacuum with a base pressure lower than 5 $\times$ 10$^{-11}$ Torr.

\subsection{DFT calculations}

All DFT calculations were performed with Vienna ab initio simulation package (VASP) code~\cite{Kresse1996a, Kresse1999},  with Perdew-Burke-Ernzerhof exchange-correlation functional~\cite{Perdew1996a}. The energy cutoff of plane wave basis is 450 eV, and 3D Brillouin zone is sampled with k-point mesh of $6\times6\times6$. All atoms are relaxed until residual force is under 0.01 eV/\AA. A tight-binding model of 12 orbitals is fitted from DFT results with Wannier functions, as implemented in Wannier90 package~\cite{pizzi2020}.

\subsection{Slave-spin calculations}
We use the $U(1)$ slave-spin method~\cite{Yu2012,Chen2022} to understand the correlation effect in \cvs. We consider a multiorbital Hubbard model read as
\begin{equation}
    H = H_0 +H_{int},
\end{equation}
where
\begin{equation}
    H_0 = \sum_{ij\alpha\beta\sigma} t_{ij}^{\alpha\beta} d_{i\alpha \sigma}^{\dagger}d_{j \beta \sigma} + \sum_{i \alpha \sigma} (\epsilon_{\alpha}-\mu)d_{i\alpha \sigma}^{\dagger} d_{i\alpha \sigma}
\end{equation}
is the tight-binding model of 12 orbitals fitted from the DFT results. $d^{\dagger}_{i\alpha\sigma}$ is the creation operator in the $i$ unit cell, with $\sigma$ denoting the spin and $\alpha=(o,s)$ enumerating both the orbital and sublattice indices respectively. In the calculation, the chemical potential $\mu$ is varied to fix the total filling in each unit cell to be $6$. The sum of the coherent and incoherent part below the chemical potential equals to 1.5 electrons per site. For the interaction part, we consider the following Hamiltonian

\begin{equation}
\begin{aligned}
    H_{int} &= \sum_{i,s} \bigg[ \frac{U}{2} \sum_{o\sigma} n_{ios\sigma} n_{ios\bar{\sigma} } + \sum_{o<o',\sigma} \big[ U' n_{ios\sigma} n_{io's\bar{\sigma} } \\
    & + (U'-J)n_{ios\sigma}n_{io's\sigma} - J(d^{\dagger}_{ios\sigma}d_{ios\bar{\sigma}}d_{io's\bar{\sigma} }^{\dagger}d_{io's\sigma} - d^{\dagger}_{ios\sigma}d^{\dagger}_{ios\bar{\sigma}} d_{io's\sigma} d_{io's\bar{\sigma} }) \big]\bigg],
\end{aligned}
\end{equation}
where $n = d^{\dagger}d$ is the density operator. $U$, $U'$ and $J$ denote the intraorbital Hubbard interaction,  interorbital repulsion and the Hund's coupling respectively. During the simulation, we take $U'=U-2J$. In the slave spin method, the electron creation operator $d^{\dagger}_{i\alpha\sigma} = S^{+}_{i\alpha\sigma}f^{\dagger}_{i\alpha\sigma}$, where slave spin $S^{+}$ represents the charge degree of freedom while the spinon operator $f^{\dagger}$ carries the spin degree of freedom. The band renormalization effect is signaled by the decreasing of the orbital resolved quasiparticle weight $Z_{\alpha} = |S^{+}_{\alpha}|^2$. The density of state is calculated by the integral over the single-electron spectral function with $A(\omega) = \sum_{k} A(k,\omega)$, where the single-electron spectral function $A(k, \omega)$ is obtained from the convolution of slave spin and slave fermion Green's functions~\cite{Chen2022}. 


\section{Acknowledgments}
This research used resources of the Stanford Synchrotron Radiation Lightsource, SLAC National Accelerator Laboratory, which is supported by the U.S. Department Of Energy (DOE), Office of Science, Office of Basic Energy Sciences under Contract No. DE-AC02-76SF00515. The ARPES work at Rice University was supported by the Gordon and Betty Moore Foundation's EPiQS Initiative through grant No. GBMF9470 (J.H.) and the Robert A. Welch Foundation Grant No. C-2175 (M.Y.). Y.Z. is partially supported by the AFOSR Grant No. FA9550-21-1-0343.
The theory work at Rice is primarily supported by the U.S. DOE, BES, under Award No. DE-SC0018197 (model building and microscopic calculation, L.C.), by the AFOSR under Grant No. FA9550-21-1-0356 (materials search, C.S.), by the Robert A. Welch Foundation Grant No. C-1411 (model conceptualization, Q.S.), and by the Vannevar Bush Faculty Fellowship ONR-VB N00014-23-1-2870 (conceptualization, Q.S.). The single-crystal synthesis work at Rice was supported by US NSF-DMR-2100741 (P.D.) and by the Robert A. Welch Foundation under grant no. C-1839 (P.D.). M.H. and D.L. acknowledge the support of the U.S. Department of Energy, Office of Science, Office of Basic Energy Sciences, Division of Material Sciences and Engineering, under contract DE-AC02-76SF00515.

\section{Author contributions}
MY oversaw the project. QS and CS first proposed the compound associated with the pyrochlore lattice. Single crystals were synthesized by BG under the guidance of PD. JH, YZ, and MY carried out the ARPES measurements with the help of DL, MH, TY and EV. The ARPES data were analyzed by JH. $U$(1) slave-spin calculations were carried out by LC, CS and QS. Density-functional theory calculations and tight-binding model fitting were carried out by YH under the guidance of BY. Transport and heat capacity measurements were carried out by YS, ZL and JC. JH and MY wrote the paper with input from all co-authors.

\section{Data Availability}
All data needed to evaluate the conclusions are present in the paper and supplementary materials. Additional data are available from the corresponding author on reasonable request.

\section{Competing interests}
The authors declare no competing interests. 

\newpage 

\end{document}